\documentclass[prl,superscriptaddress,floatfix,showpacs,aps,reprint,letterpaper,nobalancelastpage]{revtex4-1}

\usepackage[colorlinks=true]{hyperref}
\usepackage{graphicx} 
\usepackage{color} 
\definecolor{Blue}{rgb}{0.3,0.3,0.9}
\usepackage{amsthm}
\usepackage{amsmath}
\usepackage{amssymb}
\usepackage{bbm,amsfonts}
\usepackage{setspace}
\usepackage{nomencl}
\usepackage{amstext} 

\newcommand{\Csub}{\mathcal{C}}

\newcommand{\Id}{\mathbbm{1}}

\newcommand{\kett}{|\psi\rangle}

\newcommand{\Clif}{\text{Clif}_n}

\newcommand{\avg}[1]{\left\langle{#1}\right\rangle}

\newcommand{\LamC}{\Lambda_{\mathcal{C}}}
\newcommand{\LamCb}{\Lambda_{\overline{\mathcal{C}}}}
\newcommand{\LamCtil}{\Lambda_{\tilde{\mathcal{C}}}}
\newcommand{\LamCbd}{\Lambda_{\overline{\mathcal{C}},d}}
\newcommand{\pC}{p_{\mathcal{C}}}

\newcommand{\pcb}{p_{\overline{\mathcal{C}}}}
\newcommand{\rC}{r_{\mathcal{C}}}
\newcommand{\rCest}{r_{\mathcal{C}}^{\mathrm{est}}}
\newcommand{\FLamCb}{\overline{F_{\Lambda_{\overline{\mathcal{C}}},\mathcal{I}}}}
\newcommand{\FLamC}{\overline{F_{\Lambda_{\mathcal{C}},\mathcal{I}}}}
\newcommand{\FLamCtil}{\overline{F_{\Lambda_{\tilde{\mathcal{C}}},\mathcal{I}}}}
\newcommand{\FLam}{\overline{F_{\Lambda,\mathcal{I}}}}

\renewcommand{\section}[1]{{\em #1}.---}

\def\>{\rangle}
\def\<{\langle}

\begin{document}

\title{Efficient measurement of quantum gate error by interleaved randomized benchmarking}

\author{Easwar Magesan}
\affiliation{Department of Applied Mathematics, University of Waterloo, Waterloo, ON N2L 3G1, Canada}	
\affiliation{Institute for Quantum Computing, University of Waterloo, Waterloo, ON N2L 3G1, Canada}
\author{Jay M. Gambetta}
\affiliation{IBM T.J. Watson Research Center, Yorktown Heights, NY 10598, USA}
\author{B. R. Johnson}
\affiliation{Raytheon BBN Technologies, Cambridge, MA 02138, USA}
\author{Colm A. Ryan}
\affiliation{Raytheon BBN Technologies, Cambridge, MA 02138, USA}
\author{Jerry M. Chow}
\affiliation{IBM T.J. Watson Research Center, Yorktown Heights, NY 10598, USA}
\author{Seth T. Merkel}
\affiliation{IBM T.J. Watson Research Center, Yorktown Heights, NY 10598, USA}
\author{Marcus P. da Silva}
\affiliation{Raytheon BBN Technologies, Cambridge, MA 02138, USA}
\author{George A. Keefe}
\affiliation{IBM T.J. Watson Research Center, Yorktown Heights, NY 10598, USA}
\author{Mary B. Rothwell}
\affiliation{IBM T.J. Watson Research Center, Yorktown Heights, NY 10598, USA}
\author{Thomas A. Ohki}
\affiliation{Raytheon BBN Technologies, Cambridge, MA 02138, USA}
\author{Mark B. Ketchen}
\affiliation{IBM T.J. Watson Research Center, Yorktown Heights, NY 10598, USA}
\author{M. Steffen}
\affiliation{IBM T.J. Watson Research Center, Yorktown Heights, NY 10598, USA}

\begin{abstract}
We describe a scalable experimental protocol for estimating the average error of individual quantum computational gates. This protocol consists of interleaving random Clifford gates between the gate of interest and provides an estimate as well as theoretical bounds for the average error of the gate under test, so long as the average noise variation over all Clifford gates is small. This technique takes into account both state preparation and measurement errors and is scalable in the number of qubits. We apply this protocol to a superconducting qubit system and find a bounded average error of $0.003 \left[0,0.016\right]$ for the single-qubit gates $X_{\pi/2}$ and $Y_{\pi/2}$. These bounded values provide better estimates of the average error than those extracted via quantum process tomography. 
\end{abstract}

\maketitle

Determining how well an operation is implemented on a quantum device is of fundamental importance in quantum information theory. Such a characterization allows a direct comparison between different architectures for computation as well as an understanding of the performance of the building blocks of a quantum computer. The standard method for characterizing a quantum operation is quantum process tomography (QPT) \cite{CN,PCZ} which is subject to two significant drawbacks: first, it is not scalable in the number of sub-systems (qubits) comprising the system; and second if state-preparation and measurement (SPAM) errors are present, then these errors will contribute to those of the gate being characterized, hence giving an unfaithful estimation of the actual error. In many cases, one does not require the complete knowledge that QPT aims to provide. As a result, various methods for partially characterizing a quantum gate have been developed \cite{DCEL,EAZ,LLEC,ESMR,SMKE,BPP,SLCP,FL,MGE}. Ideally such a method should be scalable in the number of qubits, $n$, comprising the system as well as provide a faithful measure of the noise that is independent of SPAM errors.

One particular method for partial noise characterization is ``randomized benchmarking" (RB) \cite{EAZ,KLRB, MGE}, with Ref.~\cite{MGE} providing the first scalable RB protocol that satisfies all of the above criteria. The general idea of RB is to implement random sequences of gates that compose to the identity operation, and measure the fidelity of each sequence. Averaging over different realizations results in a fidelity decay versus the sequence length, from which the average error over the full gate set is estimated via fitting the curve to a derived model. The simplicity of this protocol has lead to various experimental implementations of the single-qubit gate protocol presented in Ref.~\cite{KLRB}, including in atomic ions with different types of traps \cite{KLRB,Biercuk2009,BWC}, liquid state nuclear spins \cite{RLL}, superconducting qubits \cite{CGT,Chow2010a,Paik2011}, and atoms in optical lattices \cite{Olmschenk2010}. 

The multi-qubit RB protocol described in Ref.~\cite{MGE} is restricted to benchmark only the full Clifford group on $n$ qubits, $\Clif$. While this provides a significant step towards scalable benchmarking of a quantum information processor it is desirable in many cases to benchmark \emph{individual} gates in $\Clif$ rather than the entire set. One method for characterizing the fidelity of single Clifford gates has been provided in Ref.~\cite{MSRL}, proposing an extension of the protocol introduced in Ref.~\cite{ESMR}. The main drawback of this method is that it does not account for SPAM errors which can bias estimates of the gate error. Note that benchmarking Clifford gates rather than general elements of the unitary group is not a significant restriction as any unitary gate can be implemented with fault-tolerance using special input states, Clifford elements and computational basis measurements~\cite{BK05}. Additionally, the unitary group can be generated via $\Clif$ through the addition of a single gate not in the group~\cite{Sho96}. Thus, benchmarking Clifford elements provides signicant information regarding the reliability of a general quantum gate, and is a relevant metric for fault-tolerant thresholds  \cite{Sho96,AB-O,KLZ,Pre97}. 

In this Letter, we present a new protocol for benchmarking individual Clifford gates via randomization. Our protocol consists of interleaving random gates between the gate, $\mathcal{C}$, of interest. In the limits of either perfect random gates or that the average error of all gates is depolarizing, our protocol estimates the gate error of $\mathcal{C}$ perfectly. In the completely general case where the random gates have arbitrary errors with small average variation, we provide explicit bounds for the error of $\mathcal{C}$. These bounds give direct information regarding the quality of computational gates and thus useful information about reaching thresholds for fault-tolerant quantum computation \cite{Sho96,AB-O,KLZ,Pre97}. The method utilizes many of the techniques of Ref.~\cite{MGE} and thus is both scalable (with time-complexity $O(n^4)$) and independent of SPAM errors. Finally, we experimentally demonstrate this protocol on a superconducting qubit, extracting a gate error of $0.003$ with theoretical lower/upper bound of $[0,0.016]$ for both $X_{\pi/2}$ and $Y_{\pi/2}$ gates ($U_{\theta}$ is a rotation of $\theta$ around axis $U$). This error is smaller than the gate errors extracted via QPT ($0.011^{+0.011}_{-0.009}$ and $0.020^{+0.009}_{-0.008}$, respectively). A similar technique was recently employed to benchmark two qubits gates in an ion trap system ~\cite{Gaebler}.  Here we provide a general expression for individual gate error for an arbitrary number of qubits.


\begin{figure}[t!]
\centering
\includegraphics[width=0.40\textwidth]{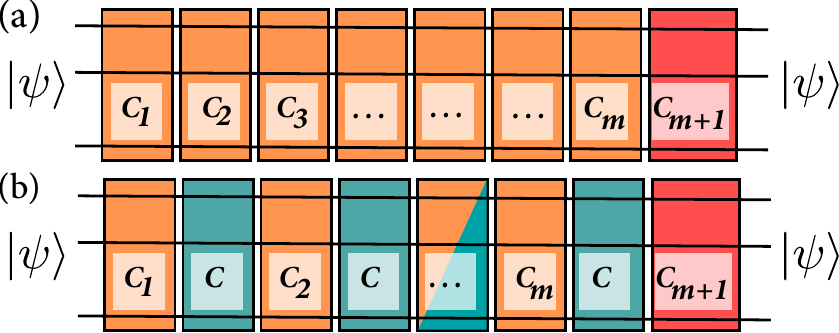}
\caption{\label{fig:1} (color online)~Randomized benchmarking protocols. (a)-(b)~Schemes for the standard and interleaved benchmarking protocols. The target gate, $\mathcal{C}$ (green) is interleaved with random gates $\mathcal{C}_{i}$ (orange) chosen from $\Clif$. A final gate $\mathcal{C}_{m+1}$ (red) is performed to make the total sequence equal to the identity operation. }
\end{figure}

\section{Interleaving benchmarking protocol} To benchmark the Clifford element $\mathcal{C}$, which has an associated noise operator $\Lambda_{\mathcal{C}}$, we fix an initial state $\kett$ and perform the following steps: 

\underline{Step 1}: Implement standard randomized benchmarking [see Fig.~\ref{fig:1}(a)] which, for completeness, we briefly summarize here (additional details in Ref.~\cite{MGE,MGE1}). For various values of $m$, 
choose $K$ sequences of random gates where the first $m$ gates are chosen uniformly at random from $\Clif$. 
The $(m+1)$th gate is chosen to be the inverse of the composition of the first $m$ random gates and can be found efficiently by the Gottesman-Knill theorem~\cite{Got97}. 
Assuming each Clifford element $\mathcal{C}_{i_j}$ for each step $j$ has some associated error, $\Lambda_{i_{j}}$, the sequence of gates is modeled by
\begin{equation}\label{eq:seqRB}
\mathcal{S}_\mathbf{i_m} = \Lambda_{i_{m+1}}\circ\mathcal{C}_{i_{m+1}}\circ \left(\bigcirc_{j=1}^{m}\left[\Lambda_{i_{j}}\circ\mathcal{C}_{i_j}\right]\right),
\end{equation}
\noindent where $\circ$ is a composition, $\mathbf{i_m}$ is the $m$-tuple $(i_1,...,i_m)$ and $i_{m+1}$ is uniquely determined by $\mathbf{i_m}$. Next, measure the probability that the initial state is not changed by the sequence,
$\mathrm{Tr} [ E_\psi \mathcal{S}_\mathbf{i_m} (\rho_\psi)]$, 
which we call the ``survival probability". Here $\rho_\psi$ is a quantum state that takes into account state-preparation errors and $E_\psi$ is the positive operator valued measure element that takes into account measurement errors.  In the ideal (noise-free) case the survival probability will be 1 for each sequence. Averaging the survival probability over the $K$ sequences gives the sequence fidelity $F_\mathrm{seq}(m,\psi)$ and a fit 
to either the zeroth or first  order model:
\begin{equation}\label{eq:model}
\begin{split}F^{(0)}_\mathrm{seq}(m,\psi)=&A_0 p^{m} + B_0,\\
F^{(1)}_\mathrm{seq}(m,\psi) =&A_1 p^{m} +C_1  (m-1) p^{m-2}+B_1,
\end{split}
\end{equation} gives the depolarizing parameter $p$ (the average error rate over all Clifford gates is given by $r = (d-1)(1-p)/d$), where $d=2^n$ is the dimension of the system. The coefficients $A_{1(0)}$ , $B_{1(0)}$ , and $C_1$ absorb the state preparation and measurement errors as well as the error on the final gate.

\underline{Step 2}: Choose $K$ sequences of Clifford elements where the first Clifford $\Csub_{i_1}$ in each sequence is chosen uniformly at random from $\Clif$, the second is always chosen to be $\mathcal{C}$, and alternate between uniformly random Clifford elements and deterministic $\mathcal{C}$ up to the $m$th random gate [see Fig. \ref{fig:1}(b)]. The $(m+1)$th gate is chosen to be the inverse of the composition of the first $m$ random gates and $m$ interlaced $\mathcal{C}$ gates (we adopt the convention of defining the length of a sequence by the number of random gates). The superoperator representing the sequence is
\begin{equation}\label{eq:seq}
\mathcal{V}_\mathbf{i_m} = \Lambda_{i_{m+1}}\circ\mathcal{C}_{i_{m+1}}\circ\left(\bigcirc_{j=1}^{m}\left[\mathcal{C}\circ\Lambda_{\mathcal{C}}\circ\Lambda_{i_{j}}\circ\mathcal{C}_{i_j}\right]\right).
\end{equation} For each of the $K$ sequences, measure the survival probability $\mathrm{Tr} [ E_\psi \mathcal{V}_\mathbf{i_m} (\rho_\psi)]$ and average over the $K$ random sequences to find the new sequence fidelity
$F_{\overline{\mathrm{seq}}}(m,\psi)$.
Fit $F_{\overline{\mathrm{seq}}}(m,\psi)$ to one of the new zeroth or first order models to obtain the depolarizing parameter $\pcb$. The expressions for these models are given by Eq. \eqref{eq:model} where $p$ is 
replaced by the new depolarizing parameter $\pcb$.

\underline{Step 3}: From the values obtained for $p$ (Step 1) and $\pcb$ (Step 2), the gate error of  $\LamC$ (which is exactly given by $r_{\mathcal{C}} = 1-$average gate fidelity of $\LamC$) is estimated by
\begin{equation} \label{eq:rest}
\rCest = \frac{(d-1)\left(1-\pcb/p\right)}{d},
\end{equation}
and must lie in the range $[\rCest - E,\rCest + E]$ where 
\begin{equation}\label{eq:Error}
E = \mathrm{min}  \left\{
\begin{split}
	&\frac{(d-1)\left[\left|p-\pcb/p\right| + (1-p)\right]}{d}\\
	&\frac{2(d^2-1)(1-p)}{p d^2}+\frac{4\sqrt{1-p}\sqrt{d^2-1}}{p}.
\end{split}\right.
\end{equation}  
One interpretation of $E$ is that it arises from imperfect random gates. To see this, first note that in the limit of perfect random gates, $p\rightarrow 1$, $r_\mathcal{C}^\mathrm{est}$ goes to the standard error for a depolarizing channel with strength $\pC$ (equivalently $r_\mathcal{C}^\mathrm{est} \rightarrow \rC$), and $E$ goes to zero. In the more specific case of $\Lambda$ being a Pauli channel, one can replace the second possibility in Eq.~(\ref{eq:Error}) with
${2(d^2-1)(1-p)}/{pd^2}$ and when $\Lambda$ is depolarizing, $E = 0$. In the \emph{typical} case, we expect that $\Lambda$ will be close to a depolarizing channel and the above general bounds will over-estimate the gate error.

\section{Experimental implementation}
Using the new protocol, we verified the performance of two single-qubit gates on a superconducting transmon qubit. The device is similar to the one described in Ref.~\cite{Chow:2012}, but we focus
on just a single qubit with $\omega_{01}/2\pi = 5.4493\,\mathrm{GHz}$, anharmonicity of $(\omega_{12}-\omega_{01})/2\pi = -228\,\mathrm{MHz}$, and coherence times of $T_1 = 5.0\,\mathrm{\mu s}$ and $T_2^{\mathrm{echo}} = 3.2\,\mathrm{\mu s}$.

Single-qubit control was performed by means of shaped microwave pulses applied to capacitively-coupled bias lines that address individual qubits. We used Gaussian shaped pulses with a derivative envelope applied to an orthogonal quadrature to minimize errors due to higher levels of the transmon~\cite{Motzoi:2009}. The Gaussian width was $\sigma = 5\,\mathrm{ns}$ and the pulses were truncated to have a total duration of $4\sigma = 20\,\mathrm{ns}$. A pulse calibration procedure was used which employed several sequences of repeated pulses that amplify small rotation angle and phase errors. A Levenberg-Marquardt search provided all calibrated pulse parameters in a few minutes.

\begin{figure}[t!]
\centering
\includegraphics[width=0.40\textwidth]{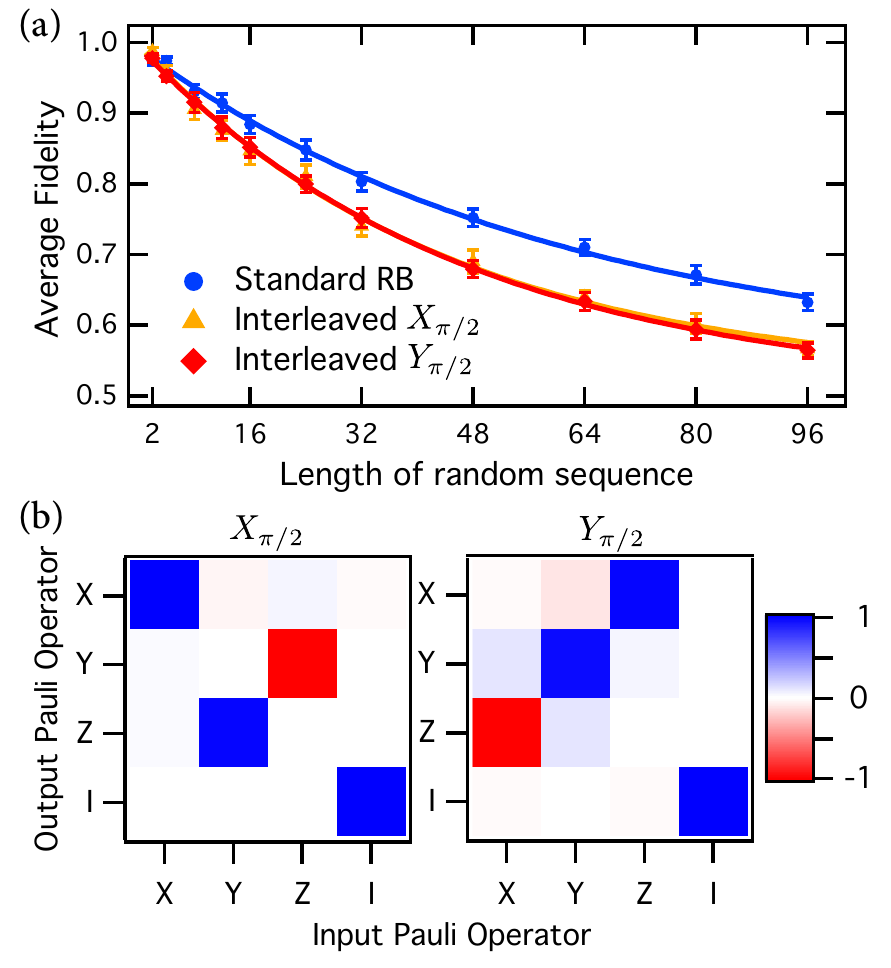}
\caption{\label{fig:2} Experimental implementation of interleaved RB. (a)~Measurement of average fidelity over 32 random sequences each of lengths between 2 and 96. The blue data (circles) show the result of the standard RB protocol, while red (triangles) and orange (diamonds) data correspond to interleaved sequences for $X_{\pi/2}$ and $Y_{\pi/2}$ gates, respectively. All data are well described by the first model of Eq.~\eqref{eq:model}, with $p = 0.984\pm0.004$ (standard RB), $\pcb = 0.978\pm0.005$ (interleaved $X_{\pi/2}$) and $\pcb = 0.979\pm0.001$ (interleaved $Y_{\pi/2}$). Error bars are the standard error of the mean of each point. (b)~Pauli transfer maps from process tomography of the $X_{\pi/2}$ and $Y_{\pi/2}$ gates with corresponding gate errors of $0.011^{+0.011}_{-0.009}$ and $0.020^{+0.009}_{-0.008}$, respectively.}
\end{figure}

To perform standard randomized benchmarking, we chose a Clifford generating set of $s = \{I, X_{\pm\pi/2}, X_{\pi}, Y_{\pm\pi/2}, Y_{\pi}$\}. Each Clifford gate in a random sequence is performed by a random choice from the set of minimal length constructions of that gate. For the generating set $s$, a Clifford gate has an average length of $1.875$ pulses. To find the average fidelity for sequences of length $N$, we create 32 random sequences of $N+1$ Clifford gates, measure $\avg{\sigma_z}$ at the end of each, and then average the results. Figure \ref{fig:2}(a) shows the measured average fidelities (blue circles) versus sequence length. The data fit well to the first model of Eq.~\eqref{eq:model} with $p = 0.984\pm0.004$, corresponding to an estimated average error rate for the entire Clifford group of $r = (1-p)/2 = 0.008\pm0.002$, which is in reasonable agreement with the expected error of 0.006 from decoherence.

Since the Clifford generating rotations in $s$ can each be implemented with just a single pulse, we expect a lower error rate for such gates than the average over the entire group. We verify this for $X_{\pi/2}$ and $Y_{\pi/2}$ gates with interleaved benchmarking.  The resulting average fidelities for the $X_{\pi/2}$ and $Y_{\pi/2}$ interleaved sequences are shown in Fig.~\ref{fig:2}(a) as orange triangles and red diamonds, respectively. The fidelities are lower than the standard RB results because of an effective doubling in the number of pulses in the interleaved sequences. These sequences fit to a model with the $\pcb = 0.978\pm0.005$ and $0.979\pm0.001$. By Eq.~\eqref{eq:rest}, this gives our best estimated error rate for $X_{\pi/2}$ and $Y_{\pi/2}$ gates of $r_\mathcal{C}^\mathrm{est} =0.003\pm 0.003$ and Eq.~\eqref{eq:Error} provides bounds of $\left[0,0.016\right]$. 

To further test the robustness of the technique, we also intentionally introduce additional error on a target gate to test the sensitivity of the interleaved benchmarking protocol to calibration errors. These results are summarized in Table~\ref{table:1}. For the small set of calibration errors introduced, the model reliably tracks the anticipated pulse infidelity.

We compare the interleaved RB result to the standard method of measuring gate performance by performing QPT. The process matrices for the $X_{\pi/2}$ and $Y_{\pi/2}$ gates are shown in Fig.~\ref{fig:2}(b) in the Pauli basis of the Liouville representation (also known as the Pauli transfer map, see \cite{Chow:2012}). To extract these maps we employ maximum likelihood estimation (MLE) to ensure that the maps are physical (we require the maps to be completely positive, but allow them to be non-trace-preserving because of potential leakage out of the qubit space). The gate errors extracted from these maps are $1-\mathcal{F} =  0.011^{+0.011}_{-0.009}$ and $0.020^{+0.009}_{-0.008}$ respectively. We attribute the increase in error seen in QPT to SPAM errors.  Additionally, the use of MLE leads to difficulties in assigning error bars to the fidelities through Monte-Carlo bootstrapping.
Consequently, interleaved benchmarking provides a more reliable estimate for the performance of Clifford gates.

\begin{table}
\begin{ruledtabular}
\begin{tabular}{cccc}
    Amp. error $\epsilon$ & $r_\mathrm{th}$ & $\rCest$ & $\rCest$ bound\\
    \hline
    0.0  & 0.000 & $0.003\pm0.003$ & $\left[0,0.016\right]$\\
    $\pi/20$ & 0.004 & $0.011 \pm 0.004$ & $\left[0,0.022\right]$\\
   $\pi/10$ & 0.016 & $0.029 \pm 0.008$& $\left[0,0.058\right]$\\
    \end{tabular}
\end{ruledtabular}
 \caption{\label{table:1} Gate errors extracted with interleaved RB for intentional pulse miscalibration errors of $X_{\pi/2}$. The first column is the applied over-rotation about the $X$ axis with predicted $\Lambda_\mathcal{C}=\exp[-i \epsilon \sigma_x /2]$, the second column is found using $r_\mathrm{th}=2(1-\cos^2(\epsilon/2))/3$, the third is the experimentally extracted gate errors via interleaving with fit uncertainties, and the fourth is the bounds from Eq.~\eqref{eq:Error}.}
\end{table}

\section{Derivation of the fitting models, gate errors and bounds}
The main idea behind the derivation of the fitting models is the following ``unitary 2-design" property of the Clifford group: If $\Lambda$ is a quantum channel and $\Clif = \{\mathcal{C}\}$ then the ``twirl" of $\Lambda$, $\mathcal{W}(\Lambda)$, defined by
\begin{eqnarray}
\mathcal{W}(\Lambda)&:=& \frac{1}{\left|\Clif\right|}\sum _{j=1}^{\left|\Clif\right|} \mathcal{C}_j\circ \Lambda \circ \mathcal{C}_j^{\dagger}
\end{eqnarray}
\noindent is the unique depolarizing channel $\Lambda_{d}$ with the same average fidelity as $\Lambda$ \cite{DCEL}. The average fidelity of $\Lambda$ is given by
\begin{gather}
\overline{\mathcal{F}_{\Lambda,\mathcal{I}}} :=\overline{ \text{tr}\left(|\phi\rangle \langle \phi | \Lambda (|\phi\rangle \langle \phi |)\right)} \label{eq:avgfidelity}
\end{gather}
\noindent which is just the average over all pure states $|\phi\rangle\langle \phi |$ of the usual fidelity between the output state $\Lambda \left(|\phi\rangle \langle \phi |)\right)$ and input state $|\phi\rangle\langle \phi |$. Hence if $\Lambda_{d}$ is given by $\Lambda_{d}(\rho)=p\rho + (1-p)\frac{\Id}{d}$ then
$\overline{\mathcal{F}_{\Lambda,\mathcal{I}}}=p+\frac{(1-p)}{d}$. 

We now provide a brief overview of the derivation of the fitting models. Defining  $
\mathcal{D}_{i_j} = \mathcal{C}_{i_j} \circ \bigcirc_{s=1}^{j-1} \left[ \mathcal{C}\circ\mathcal{C}_{i_s}\right] $
allows us to write the interleaving sequence as 
\begin{gather}
\mathcal{V}_{\bf{i_m}} =\left( \Lambda_{i_{m+1}}\right) \circ \left(\bigcirc_{j=1}^m\left[{\mathcal{D}_{i_j}}^{\dagger} \circ {\Lambda}_{\mathcal{C}} \circ \Lambda_{i_j} \circ {\mathcal{D}_{i_{j}}}\right] \right).\label{eq:changeofvar}
\end{gather} The zeroth order model corresponds to the noise being independent of the gate, i.e. $\Lambda_{i_j}=\Lambda$ is independent of $\mathcal{D}_{i_j}$ for every $j$. In this case when we average over many sequences in Eq.(\ref{eq:changeofvar}) we obtain a composition of twirls of ${\Lambda}_{\mathcal{C}} \circ \Lambda$. Hence 
we obtain a composition of depolarizing channels $\LamCbd = \left({\Lambda}_{\mathcal{C}} \circ \Lambda\right)_d$ where for any state $\rho$,
$\LamCbd(\rho)=\pcb\rho + (1-\pcb)\frac{\openone}{d}$.
 Here, $1-\pcb$ corresponds to the depolarizing strength of ${\Lambda}_{\mathcal{C}} \circ \Lambda$. 

The first order model corresponds to the case where the noise depends on the gate. In this case, we apply a perturbative argument similar in nature to that of Ref. \cite{MGE} (for more details see Ref. \cite{MGE1}) to derive the fitting model. Each $\Lambda_{i}$ is perturbed about the average of all the $\Lambda_i$, denoted by $\Lambda$, and provided 
the \emph{average variation} of the $\|\delta \Lambda_i\|$, $\gamma:=\frac{1}{\Clif}\sum_i \|\delta \Lambda_i\|$, is small (ie. $\gamma^2 \ll 2/[m(m+1)]$) the first order model is a valid description of the fidelity decay curve. Note that the norm $\| \cdot \|$ can be any norm satisfying certain general properties (see Ref~\cite{MGE1} for more detail). One usually chooses the weakest norm satisfying these properties which allows for the largest class of gate-dependent errors. It is important to emphasize that the type of the noise is irrelevant for this sufficient condition, as long as the average of the magnitudes is sufficiently small, the analysis can be terminated at first order.

We now outline how to obtain the expression for the error given by Eq. (\ref{eq:rest}) as well as the various expressions for $E$ given by Eq.'s \eqref{eq:Error} and in the surrounding text. Let us begin by looking at the difference in average fidelity between $\LamCb={\Lambda}_{\mathcal{C}} \circ \Lambda$ and $\LamCtil:={\Lambda}_{\mathcal{C}} \circ \Lambda_d$, $\left|\FLamCb-\FLamCtil\right|$. Since $\Lambda_d$ is depolarizing,
\begin{gather}
\left|\FLamCb-\FLamCtil\right| = \left|\FLamCb-p\FLamC-\frac{1-p}{d}\right|. \label{eq:absvalfid}
\end{gather}
\noindent 
Hence,
\begin{equation}
\frac{\FLamCb}{p}-\frac{1-p}{dp}-E \leq \FLamC \leq \frac{\FLamCb}{p}-\frac{1-p}{dp}+E
\end{equation}
where $E$ is an upper bound for $\left|\FLamCb-\FLamCtil\right|/p $. Using $\rC :=1-\FLamC$ and $\FLamCb = \pcb+\frac{(1-\pcb)}{d}$ we find Eq. (\ref{eq:rest}). 

The first expression in Eq. (\ref{eq:Error}) can be obtained by noting that $\left|\FLamCb-\FLamCtil\right|$ can be upper bounded by
\begin{equation}
\begin{split}
\: \: \: & \left| \FLamCb-p\FLam - \frac{(1-p)}{d}\right|+ p\left|\FLam-\FLamC\right| \\ \leq & \frac{(d-1)\left[\left|\pcb - p^2 \right| + \left(p-p^2\right)\right]}{d}	
\end{split}
\end{equation}
where we have used $\left|\FLam-\FLamC\right| \leq \frac{(d-1)(1-p)}{d}$.

The second expression in Eq. (\ref{eq:Error}) is obtained by first noting that,
\begin{gather}
\left|\FLamCb-\FLamCtil\right| \leq\|\Lambda-\Lambda_d\|_{\diamond}, \label{eq:normbound}
\end{gather}
where $\| \: \|_{\diamond}$ is the ``diamond norm"~\cite{Kit97}. By the triangle inequality,
\begin{gather}
\left|\FLamCb-\FLamCtil\right| \leq  \|\Lambda-\mathcal{I}\|_{\diamond}  + \frac{2(d^2-1)(1-p)}{d^2}, \label{eq:normbound1}
\end{gather}
\noindent where $\|\Lambda_d -\mathcal{I}\|_{\diamond} = \frac{2(d^2-1)(1-p)}{d^2}$~\cite{MGE1}. It can be shown that for arbitrary $\Lambda$~\cite{SK11},
\begin{gather}
 \|\Lambda-\mathcal{I}\|_{\diamond}  \leq 4\sqrt{1-p}\sqrt{d^2-1}, \label{eq:normbound3}
\end{gather}
which gives the second expression in Eq.~(\ref{eq:Error}).

In the case of $\Lambda$ being equal to a Pauli channel,
$
\|\Lambda-\mathcal{I}\|_{\diamond} = 2(d^2-1)(1-p)/d^2
$
 always holds~\cite{MGE1}.
Lastly, the depolarizing case is obtained by noting that $\LamCb=\LamCtil$ and as such $E=0$ by definition. 

%

\section{Conclusion}
We have presented a scalable protocol for benchmarking individual quantum gates. We explicitly derive various bounds for the error of the imperfect gate in terms of parameters that are output from the protocol. The gate error can be estimated exactly in the limit of perfect gates or if the average of the error operators over all gates is depolarizing, which we believe is close to the typical case. The method is scalable in the size of the quantum system and is independent of SPAM errors. We have applied this protocol to a superconducting qubit and shown the gate errors for each of $X_{\pi/2}$ and $Y_{\pi/2}$ rotations to be lower estimates than those obtained using QPT.

We acknowledge discussions with Antonio C\'orcoles, John Smolin and Joseph Emerson. EM acknowledges support from NSERC, CIFAR and the Ontario government. We acknowledge support from IARPA under contract W911NF-10-1-0324. All statements of fact,
opinion or conclusions contained herein are those of the
authors and should not be construed as representing the
official views or policies of the U.S. Government.


\begin{thebibliography}{10}

\bibitem{CN}
I.~Chuang and M.~Nielsen, J. Mod. Opt. \textbf{44}, 2455 (1997).

\bibitem{PCZ}
J.~F. Poyatos, J.~I. Cirac, and P.~Zoller, Phys. Rev. Lett. \textbf{78}, 390
  (1997).

\bibitem{DCEL}
C.~Dankert et~al., Phys. Rev. A \textbf{80}, 012304 (2009).

\bibitem{EAZ}
J.~Emerson, R.~Alicki, and K.~Zyczkowski, Journal of Optics B: Quantum and
  Semiclassical Optics \textbf{7}, S347 (2005).

\bibitem{LLEC}
B.~Levi et~al., Phys. Rev. A \textbf{75}, 022314 (2007).

\bibitem{ESMR}
J.~Emerson et~al., Science \textbf{317}, 1893 (2007).

\bibitem{SMKE}
M.~Silva et~al., Phys. Rev. A \textbf{78}, 012347 (2008).

\bibitem{BPP}
A.~Bendersky, F.~Pastawski, and J.~Paz, Phys. Rev. Lett. \textbf{100}, 190403
  (2008).

\bibitem{SLCP}
M.~P. da~Silva, O.~Landon-Cardinal, and D.~Poulin, Phys. Rev. Lett. \textbf{107}, 210404
 (2011).


\bibitem{FL}
S.~T. Flammia and Y.-K. Liu, Phys. Rev. Lett. \textbf{106}, 230501 (2011).

\bibitem{MGE}
E.~Magesan, J.~M. Gambetta, and J.~Emerson, Phys. Rev. Lett. \textbf{106},
  180504 (2011).

\bibitem{KLRB}
E.~Knill et~al., Physical Review A \textbf{77}, 012307 (2008).

\bibitem{Biercuk2009}
M.~J. Biercuk et~al., Quantum Inf. Comput. \textbf{9}, 0920 (2009).

\bibitem{BWC}
K.~R. Brown et~al., Phys. Rev. A \textbf{84}, 030303 (2011).

\bibitem{RLL}
C.~Ryan, M.~Laforest, and R.~Laflamme, New J. Phys. \textbf{11}, 013034 (2009).

\bibitem{CGT}
J.~M. Chow et~al., Phys. Rev. Lett. \textbf{102}, 090502 (2009).

\bibitem{Chow2010a}
J.~M. Chow et~al., Phys. Rev. A \textbf{82}, 040305 (2010).

\bibitem{Paik2011}
H. Paik et~al., Phys. Rev. Lett. \textbf{107}, 240501 (2011).

\bibitem{Olmschenk2010}
S.~Olmschenk et~al., New J. Phys. \textbf{12}, 113007 (2010).

\bibitem{MSRL}
O.~Moussa et~al., arXiv:1112.4505.

\bibitem{BK05}
S.~Bravyi and A.~Kitaev, Phys. Rev. A \textbf{71}, 022316 (2005).

\bibitem{Sho96}
P.~Shor, in \emph{Proceedings of the 37'th Annual Symposium on Foundations of
  Computer Science (FOCS)} (IEEE Press, Burlington, VT, 1996).

\bibitem{AB-O}
D.~Aharonov and M.~Ben-Or, in \emph{Proceedings of the 29th Annual ACM
  Symposium on Theory of Computing (STOC)} (1997).

\bibitem{KLZ}
E.~Knill, R.~Laflamme, and W.~Zurek, Proc. R. Soc. Lond. A \textbf{454}, 365
  (1997).

\bibitem{Pre97}
J.~Preskill, arXiv:quant-ph/9712048.

\bibitem{MGE1}
E.~Magesan, J.~M. Gambetta, and J.~Emerson, arXiv:1109.6887.

\bibitem{Got97}
D.~Gottesman, arXiv:quant-ph/9705052.

\bibitem{Koch:2007}
J.~Koch et~al., Phys. Rev. A \textbf{76}, 042319 (2007).

\bibitem{Motzoi:2009}
F.~Motzoi et~al., Phys. Rev. Lett. \textbf{103}, 110501 (2009).

\bibitem{Chow:2012}
J.~M. Chow et~al., arXiv:1202.5344.

\bibitem{Kit97}
A.~Kitaev, Russian Mathematical Surveys \textbf{52}, 1191 (1997).

\bibitem{SK11}
B.~Salman and R.~Koenig, New J. Phys. \textbf{13} (2011).

\bibitem{Gaebler}
J.~P. Gaebler et~al., arXiv:1203.3733.

\end{thebibliography}
\end{document}